# Constructing Mechanical Design Agent Based on Large Language Models

Jiaxing Lu[a], Heran Li[a], Fangwei Ning[a], Yixuan Wang[a], Xinze Li[a], and Yan, Shi[a,*]

[a] *Beihang University, 37 Xueyuan Road, Beijing, China*

***Abstract***— Since ancient times, mechanical design aids have been developed to assist human users, aimed at improving the efficiency and effectiveness of design. However, even with the widespread use of contemporary Computer-Aided Design (CAD) systems, there are still high learning costs, repetitive work, and other challenges. In recent years, the rise of Large Language Models (LLMs) has introduced new productivity opportunities to the field of mechanical design. Yet, it remains unrealistic to rely on LLMs alone to complete mechanical design tasks directly. Through a series of explorations, we propose a method for constructing a comprehensive Mechanical Design Agent (MDA) by guiding LLM learning. To verify the validity of our proposed method, we conducted a series of experiments and presented relevant cases.

## 1. Introduction

Mechanical design has evolved from hand-drawn sketches and drawing tool-assisted design to Computer-Aided Design (CAD). Currently, CAD software is widely used across various industrial fields, including aerospace, construction, medical, automotive, and digital media, to create 3D shapes. Despite the advancements, current interactive CAD design methods still have limitations. Beginners often need extensive domain-specific knowledge, and even experienced designers must invest significant time in creating and modifying design models, despite the considerable efficiency improvements CAD has brought to the process.

The emergence of Large Language Models (LLMs) exemplified by GPT-4, presents significant opportunities across various industries, including mechanical design, which stands to gain a substantial boost in productivity. Generative AI tools, such as LLMs, have the potential to address challenges in current mechanical design efforts by lowering design barriers and offering an intuitive, consistent, and user-friendly interface throughout every stage of the design process.

Our research indicates that if users directly ask LLMs to generate a part model, the quality of the resulting model can be unpredictable. Therefore, a specialized procedural approach is required to guide LLMs in learning the relevant rules. Once users complete these operations, LLMs will be able to generate high-quality part models directly in accordance with user requirements.

In this study, we demonstrate for the first time how LLMs can be utilized to develop a mechanical design agent (MDA) capable of generating high-quality part models from simple text guidance.

## 2. Related Works

To better elucidate our work, we briefly summarize the latest advancements in building generative agents based on LLMs and the sequence representation method of computer modeling.

### 2.1 Generative AI based on LLMs

Large Language Models (LLMs) have emerged as a powerful tool for generative tasks, enabling the creation of human-like text, images, and other media. These models are trained on vast amounts of data to capture the complexities of natural language, and have shown impressive performance across a wide range of applications.

Bommasani et al [1] provides an overview of the key concepts and characteristics of LLMs. It explains how these models leverage deep learning techniques to learn rich representations of language, allowing them to generate coherent and contextually-appropriate content. The authors discuss the strengths of LLMs, such as their ability to capture long-range dependencies and perform few-shot learning.

Several studies have explored ways to improve the efficiency and performance of LLM-based generative models. Sanh et al [2] and Schwartz et al [3] investigate methods to scale up the capabilities of LLMs while reducing their computational and memory requirements. This includes techniques like parameter sharing and model compression, which aim to maintain high performance while making the models more practical for real-world deployment.

The application of LLMs in specialized domains, such as healthcare, has also been an active area of research. Alsentzer et al [4] and Huang et al [5] examine the opportunities and challenges of using LLM-powered generative models for tasks like medical report generation and diagnosis support. These studies highlight the potential benefits but also the need to address issues around privacy, safety, and regulatory compliance.

Beyond text generation, researchers have explored the use of LLMs for other generative tasks. Zhavoronkov et al [6] discuss the application of LLMs to the generation of molecular structures and other complex data types, demonstrating the versatility of these models. However, these papers also acknowledge the limitations of LLMs, such as their lack of interpretability and the potential for biases and safety concerns.

The rapid progress in LLM-based generative models has also raised important questions about their evaluation and deployment. Ethayarajh et al [7] and Bender et al [8] critically examine the current state of LLM benchmarking, highlighting the need for more comprehensive and dynamic assessment frameworks that can capture the nuanced behaviors and potential risks of these models.

### 2.2 Computational Modeling Sequence Representation

Computational modeling lies at the heart of design computation, enabling us to digitally capture and adjust the core elements of design, including shape construction, connection mechanisms, and material properties. While there are various methods to express designs, we prefer those that are concise and easy to edit. Contemporary Computer-Aided Design (CAD) software defines shapes through steps such as plane drawing, stretching operations, and logical operations [9]. These steps can be implemented in simplified code via a specialized programming language (DSL), such as OnShape FeatureScript [10]. Design can also be effectively represented graphically[11, 12], where vertices represent individual components and connecting lines represent interactions between components. This approach has been efficiently and hierarchically applied to demonstrate CAD models [13], automated machines [14], advanced materials [15], building construction [16], and molecular chemistry [17]. For more complex designs, such as multi-rotor aircraft that integrate physical construction and software control, multiple DSLs can be described simultaneously. For instance, the physical construction of an aircraft can be encoded through CAD, while its control software can be encoded through a DSL specifically designed for control.

### 2.3 CAD Language Conversion

Considering the inherent sequence characteristics of the feature-driven CAD design process, a common representation method is to adopt language processing technology. Drawing inspiration from the field of Natural Language Processing (NLP) [18], several studies have focused on using language models to parse CAD sketches [19-21] while others have made similar attempts within the framework of CAD entity architecture [22, 23]. However, while these pioneering projects built generative models of actionable potential domains, they did not directly address the issue of reverse engineering. CADParser [24] takes a different approach, using Boundary-Representation (B-Rep) [25], an intermediate representation of object outlines, instead of point sets, to simplify the logical reasoning process in CAD language. The most relevant topics for our discussion are DeepCAD [26] and MultiCAD [27]. DeepSketch innovatively introduced semantically based sketch shaping equations and, in initial validation, was able to infer the CAD construction process from point group data. Building on this, MultiCAD [27] adopted a two-stage, cross-modal control learning scheme. In addition to the independent training of single modes, both [26] and [27] used forward propagation strategies to frame the boundaries of reverse engineering practice.

## 3. Methods

The objective of this research is to investigate how Large Language Models (LLMs) can be leveraged in the field of mechanical design to develop a mechanical design agent (MDA) capable of excelling in modeling tasks.

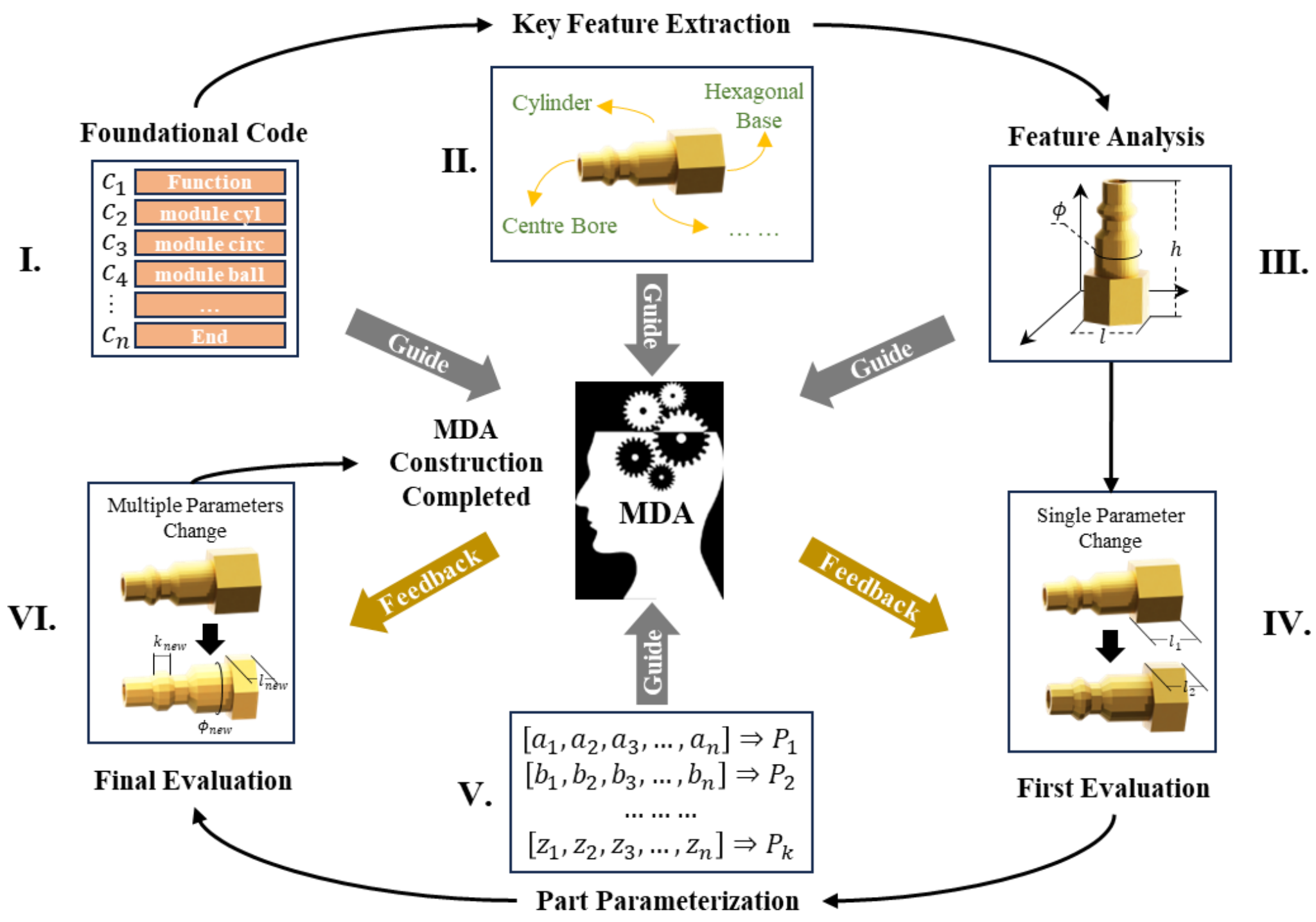

**Figure 1 Mechanical design agent (MDA) construction process.**

To achieve this, we propose a detailed construction process. Our approach is based on the concept that mechanical part models can be represented by compact programs. Essentially, mechanical design can be framed as a translation task, where natural language inputs are converted into modeling programs. LLMs are particularly well-suited for this task and offer potential enhancements to traditional solutions based on the materials provided. To thoroughly evaluate the mechanical design capabilities of LLMs across various domains, we have conducted a series of experiments demonstrating the effectiveness of our proposed approach to developing intelligent systems.

The following outlines our proposed process for developing a Mechanical Design Agent (MDA).

### 3.1 Foundational Code Learning

To address mechanical design tasks across various domains, we introduce the concept of a foundational code, which provides typical part codes:

$$Part: [C_1, C_2, \dots, C_n]$$

where $C_i$ represents the CAD history modeling sequence of the part, including actions such as sketches, extrusions, rotations, sweeps, and other commands commonly used in interactive CAD design software. $n$ is the max length of sequence.

Once the foundational code is input into LLMs, the models concentrate on interpreting the semantics and contextual logical relationships of the code. It is essential for LLMs to learn the syntax, specifications, and special symbols of the code, while also refining and synthesizing the logical connections between each code segment. After assimilating the foundational code, LLMs provide feedback to users, which may include suggesting potential modifications. Designers can then address any potential misunderstandings identified by the LLMs based on this feedback.

### 3.2 Key Feature Extraction

To ensure that Large Language Models (LLMs) effectively grasp the essential features of a component represented by base code, they must be directed to perform targeted feature extraction on the part. This process involves analyzing the component's primary structure and parameters to generate a detailed specification sheet. The designer plays a crucial role by specifying which aspects are important and outlining the desired format and content of the output. Upon receiving such guidance, the LLMs will decompose the component into fundamental geometries, such as cylinders and prisms, complete with associated dimensions. Additionally, the models will extract non-geometric parameters from standard codes, including color information, resolution factors, and material types. Ultimately, this approach will enable LLMs to deliver a more thorough analysis of the component. The user will then review this feedback to assess the accuracy of the LLMs' analysis and make any necessary corrections.

### 3.3 Feature Analysis

After learning the foundational code and extracting key features, LLMs gain an initial understanding of the part's structure and parametric representation in a high-dimensional vector space. However, to allow users to guide LLMs effectively using natural language for specialized mechanical design tasks, the models' understanding must align closely with the designers' intentions. Therefore, designers need to provide LLMs with real-world interpretations of key feature parameters and their interrelationships. This enables LLMs to comprehend higher-level characteristics of the part, such as the relationship between changes in specific parameters and operational conditions. For instance, a gear with a diameter of 10 may be associated with an operational pressure of 1 MPa. This learning process involves designers querying LLMs and refining their understanding based on the responses to ensure accuracy.

### 3.4 First Evaluation

Upon completing the first three steps, the designer will have essentially developed a mechanical design intelligent system. To initially validate the modeling capabilities of this system, the designer can perform an initial evaluation. This evaluation involves instructing the LLMs to modify a parameter in the foundational code according to specific requirements set by the designer. The LLMs will then provide the complete modified part code as feedback. The designer assesses whether the constructed intelligent system meets the basic requirements by calculating the difference X between the new code and the foundational code:

$$X = DIS[Code_F, Code_N] - 1$$

where $X$ can be $1, 2, 3, \dots, n$, represents the difference between the foundational code and the new code with $X = 0$ indicating an optimal result. $Code_F = [A_1, A_2, \dots, A_m]$, $Code_N = [B_1, B_2, \dots, B_m]$. $A_i$ and $B_i$ represent the modeling code sequence, $m$ is the max length of code.

### 3.5 Part Parameterization

Part parameterization is essential for enhancing the generalization capability of the MDA. LLMs achieve this by variabilizing the constant parameters in the foundational code under the designer's guidance. After performing the variabilization, LLMs provide the complete updated code, where variable substitutions replace the constants in the foundational code. The variability is quantified as $X = DIS[Code_1, Code_2] = Ns$, where $Ns$ represents the number of statements containing constants in the foundational code. During this process, the designer can impose additional constraints, such as maintaining $P_1 + P_2 = k$ throughout the parameter variations, where $P_1$, $P_2$ are parameters of the model and $k$ is a constant.

Upon completing part parameterization, the MDA's generalization ability is enhanced, enabling the generation of more design variants. This leads to improved design quality and greater alignment with the designer's intentions.

### 3.6 Final Evaluation

After completing the previous steps, the construction of the mechanical design intelligence system is finalized. The designer must now evaluate the performance of the agent to verify its final modeling capabilities. To facilitate this process, designers require the MDA to simultaneously modify several parameters while imposing only a few complex constraints, allowing the MDA to autonomously find the optimal values for the remaining parameters. Following this optimization process, LLMs will provide a final result. Next, designers generate a model based on the quality of the feedback results to determine if the MDA's generative capabilities meet the requirements. Any issues identified should be addressed promptly.

### 3.7 Summary

Guided by the aforementioned principles, LLMs can develop the Mechanical Design Agents (MDAs) as desired by the designers. An evaluated MDA can then generate a detailed 3D model from simple text, bypassing the need to undergo the complex prompt process again.

## 4. Experiments

The modeling quality of LLMs is directly related to the quality of the prompts provided during training. In this section, we demonstrate the uniqueness and feasibility of the aforementioned framework through modeling and dimension customization testing of various mechanical parts. Figure 2 illustrates both the successful and unsuccessful cases of using LLMs to construct 3D models of mechanical parts. All of our work is based on ChatGLM[28] and openscad[29] modeling code.

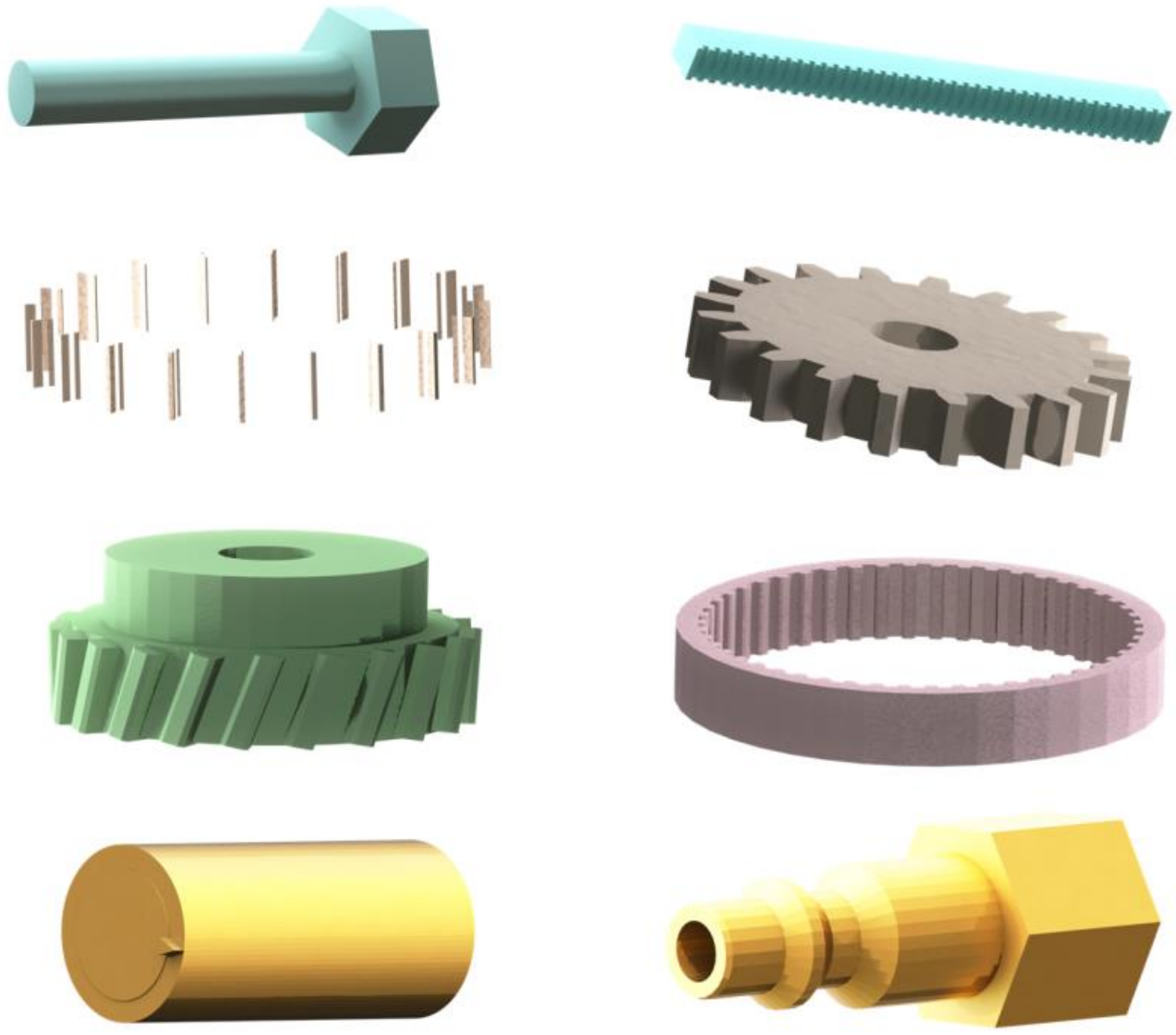

**Figure 2 Schematic diagram of mechanical part modeling and testing case based on LLMs.**

As shown in Figures 3-5, during the training of the MDA, only "one-step" is provided (merely informing the requirements for mechanical modeling). The results indicate that LLMs can still achieve their goals when constructing certain simple basic mechanical component models. However, when the models become complex, the LLMs are unable to generate the correct models.

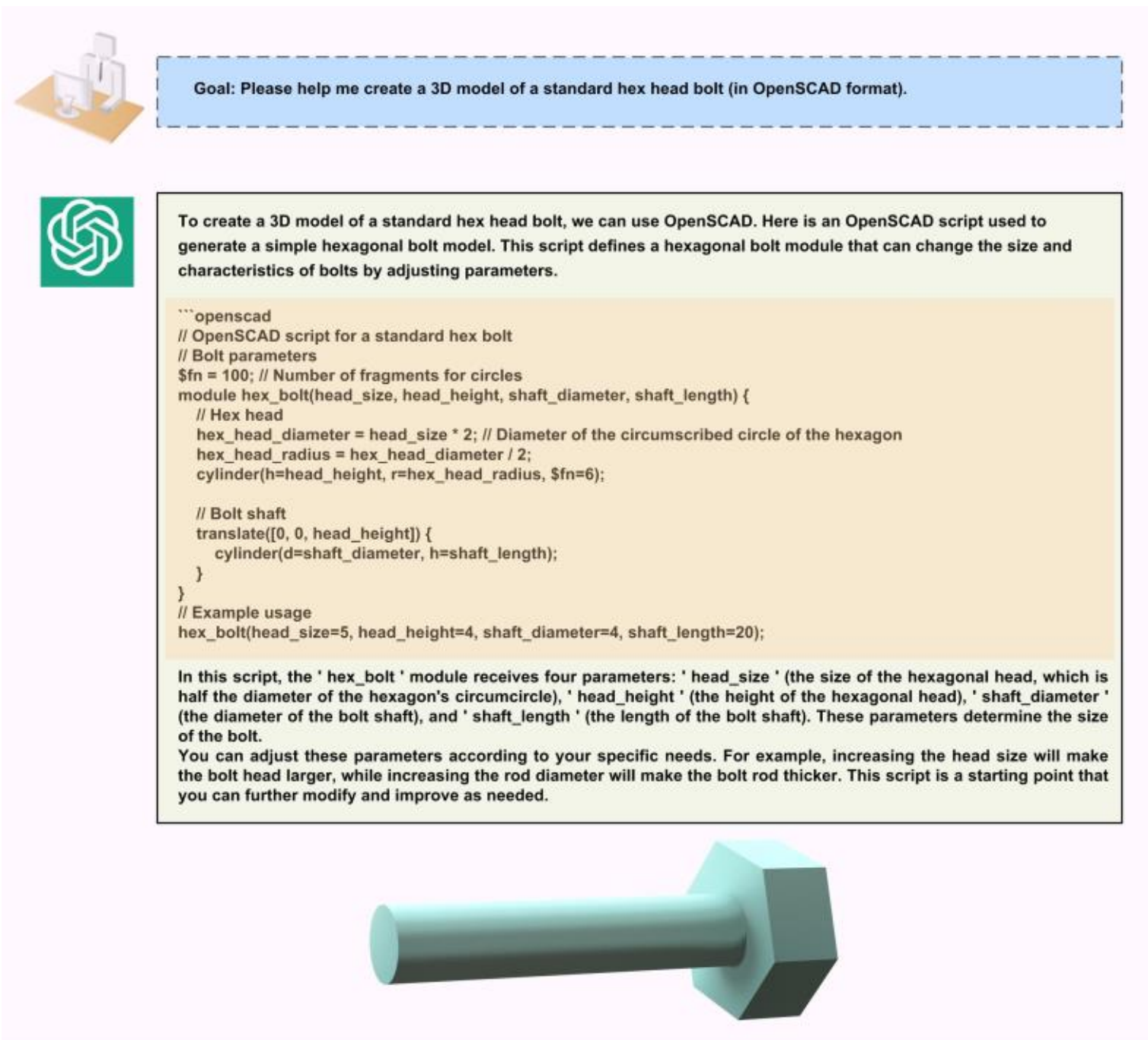

**Figure 3 Schematic diagram of LLM intelligent agent screw 3D modeling test case based on one-step prompts.**

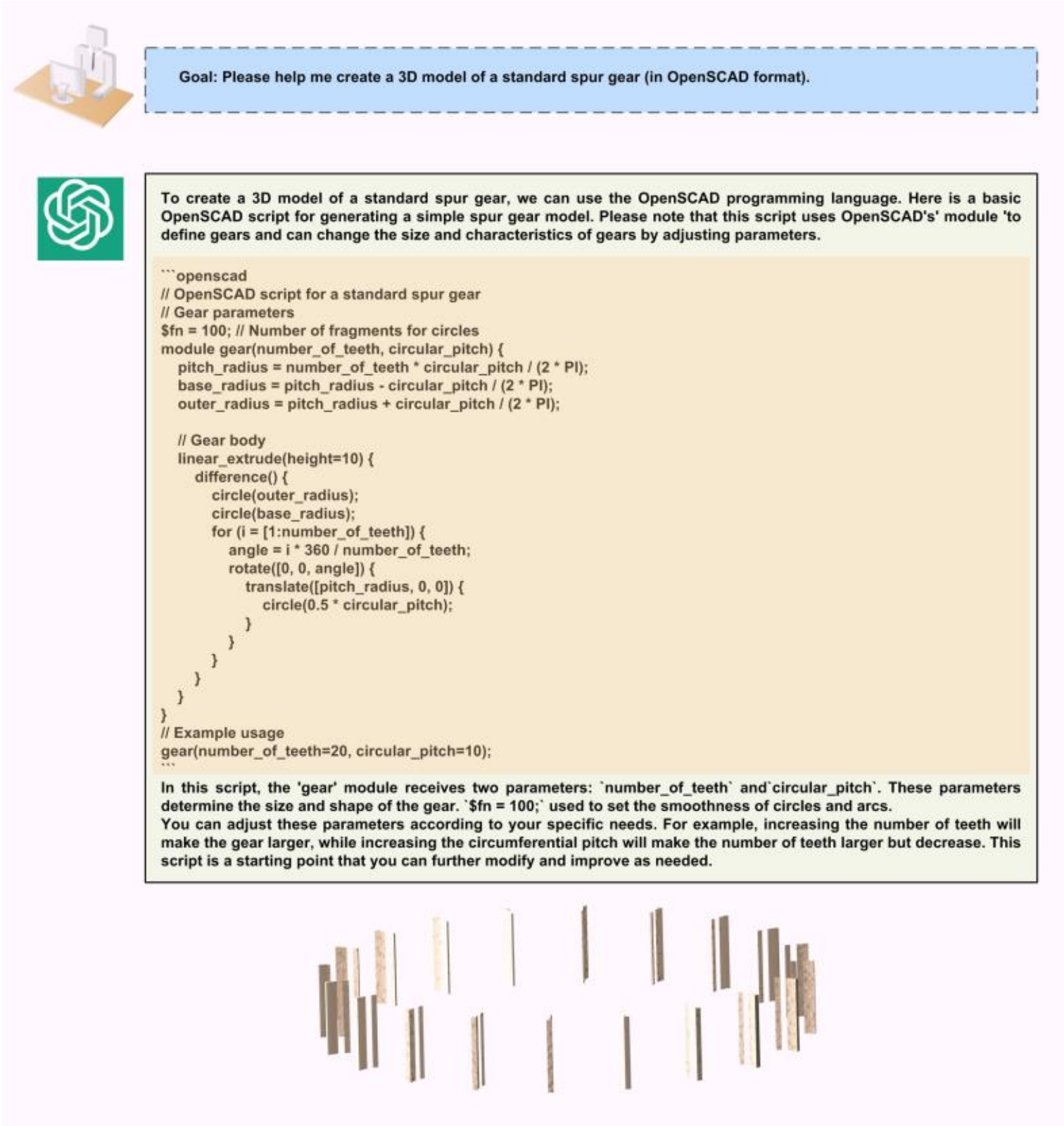

**Figure 4 Schematic diagram of LLM intelligent agent three-dimensional modeling test case of spur gear based on one-step prompt.**

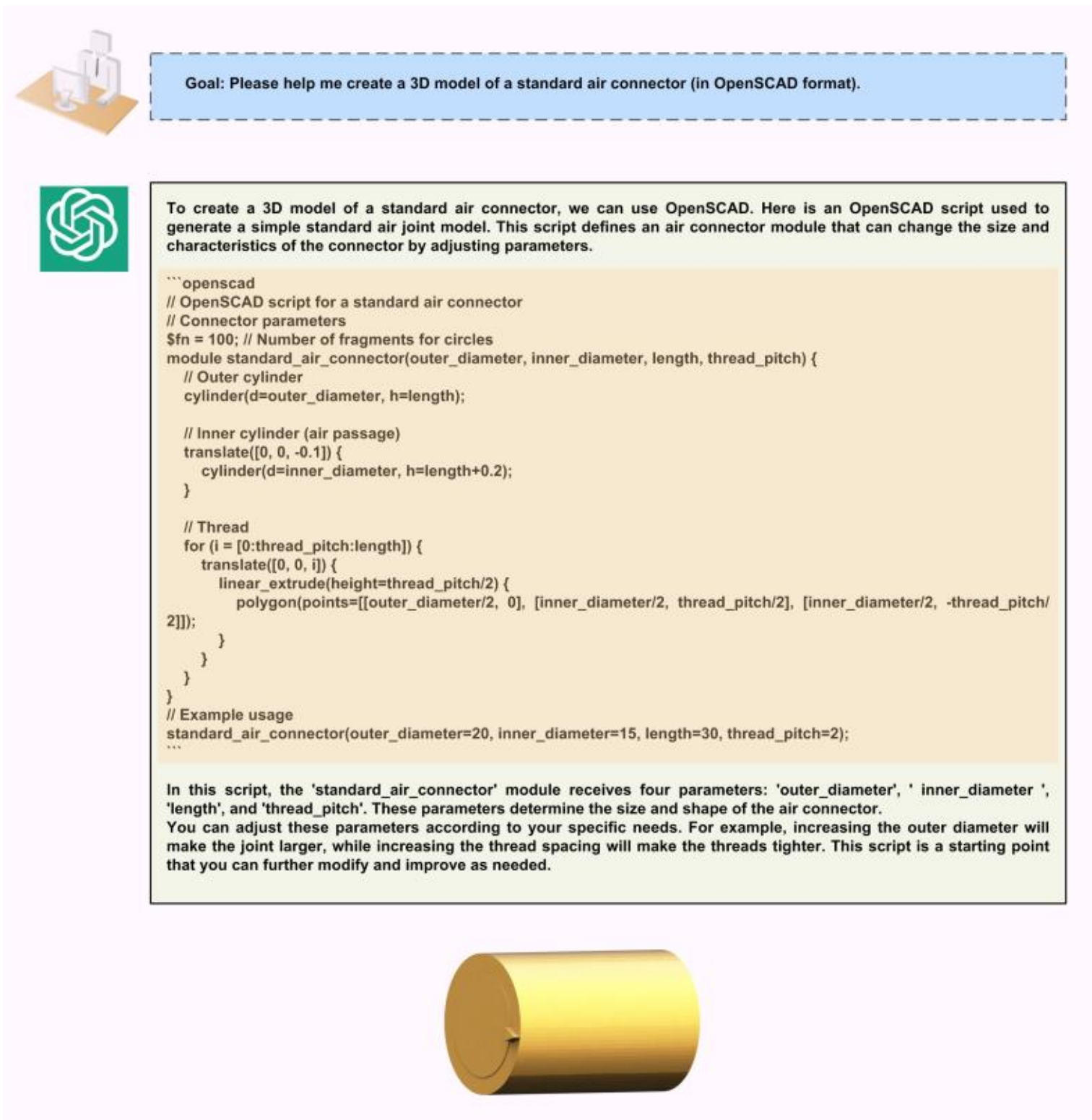

**Figure 5 Schematic diagram of LLM intelligent agent pneumatic joint 3D modeling test case based on one-step prompt.**

As shown in Figures 6-17, following the prompt framework outlined in methods section, the MDA successfully completes the high-quality modeling tasks of the rack and air fitting. Additionally, users can customize the parameter dimensions through this intelligent agent.

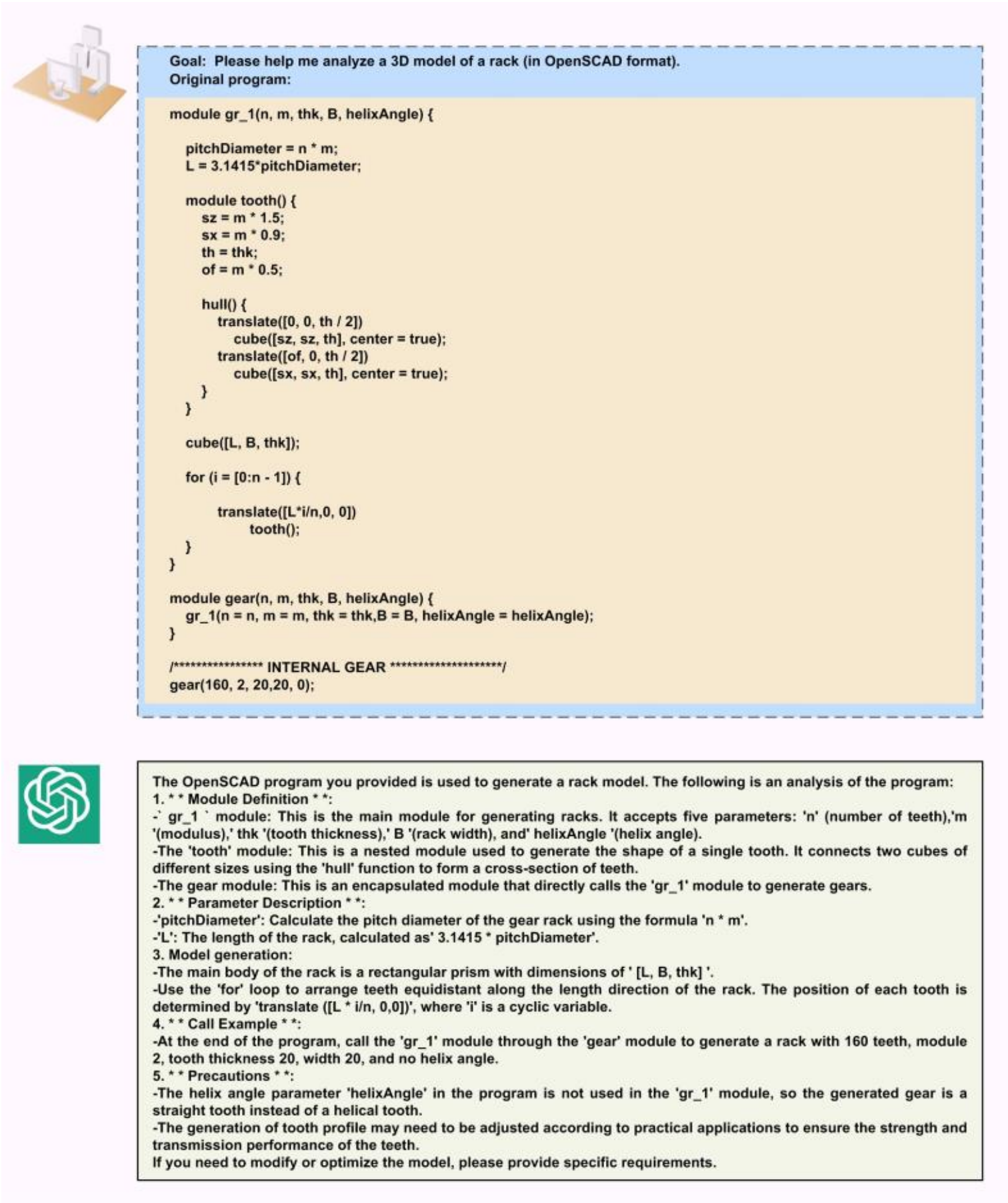

**Figure 6 Spline MDA construction process—Foundational Code Learning (Example 1).**

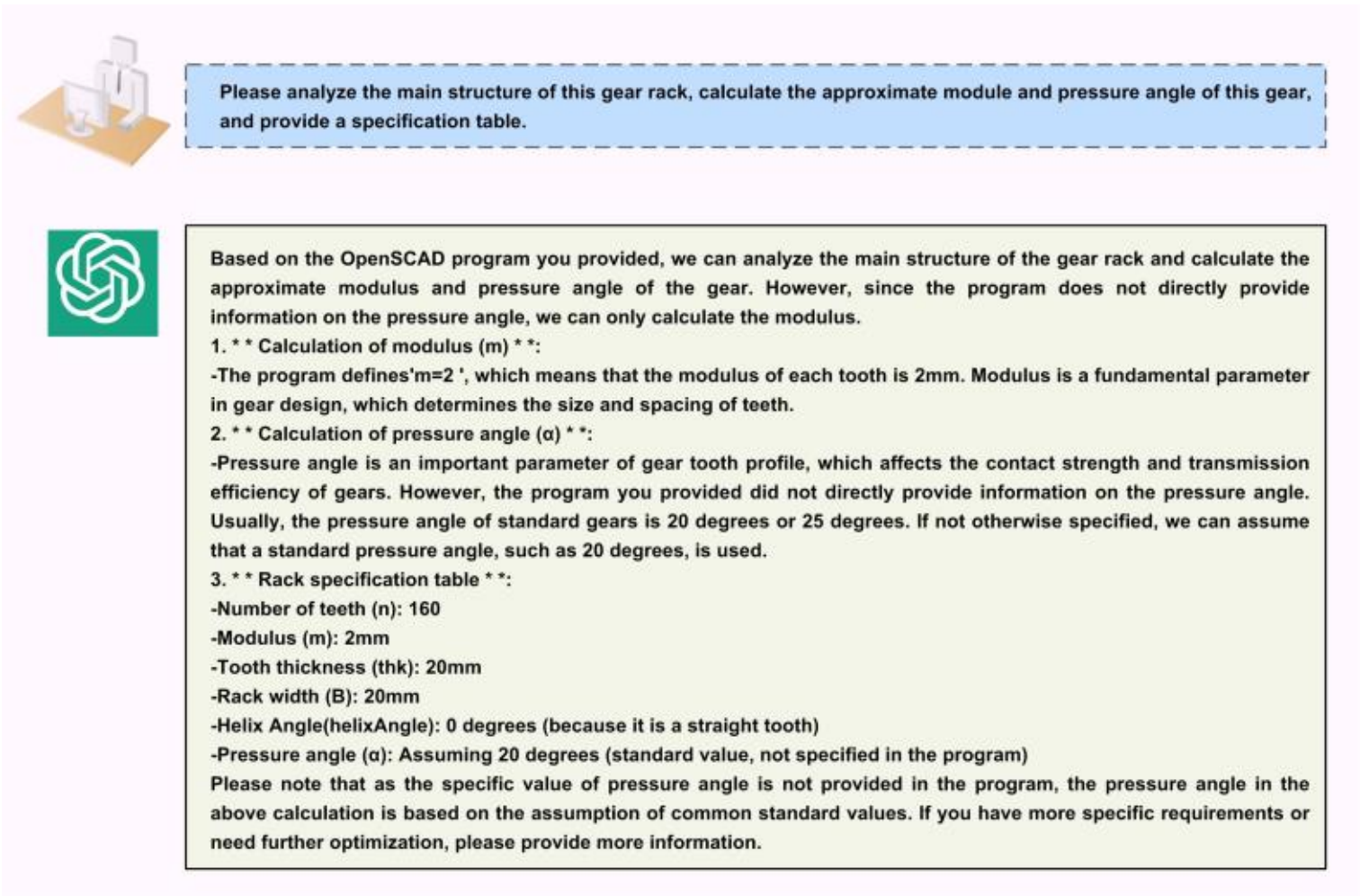

**Figure 7 Spline MDA construction process—Key Feature Extraction and Feature Analysis (Example 1).**

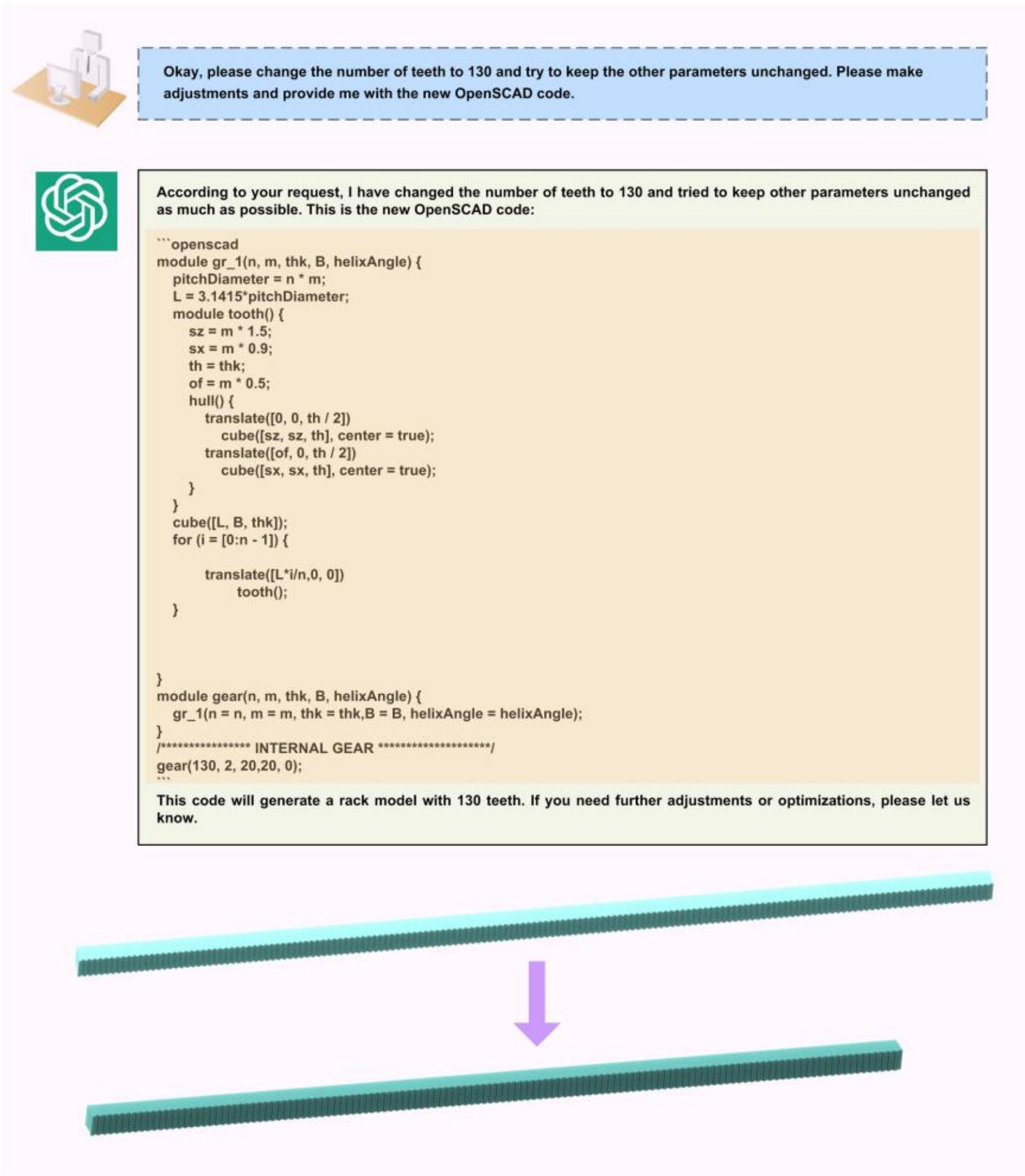

**Figure 8 MDA construction process—First Evaluation (Example 1).**

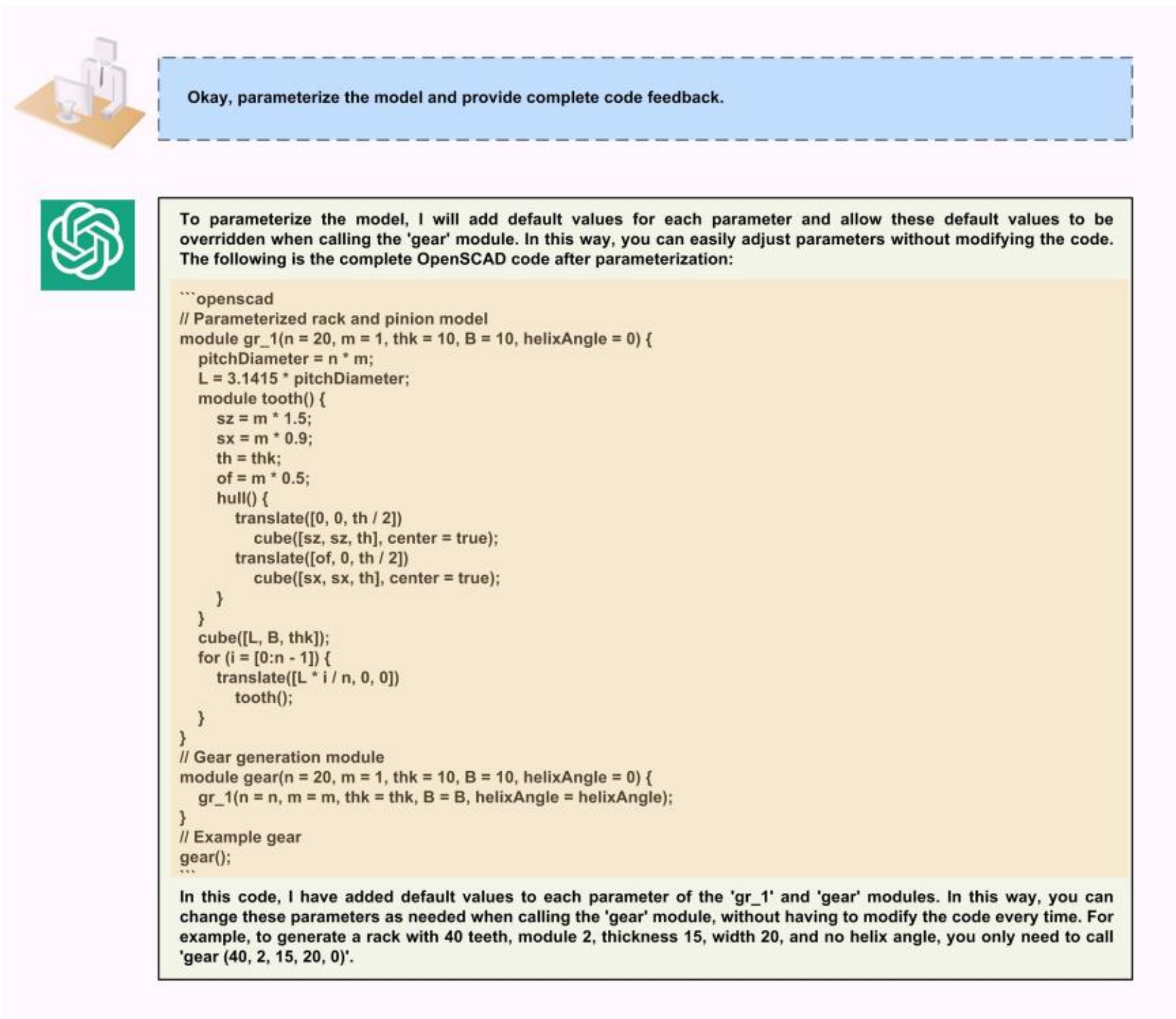

**Figure 9 Spline MDA construction process—Part Parameterization (Example 1).**

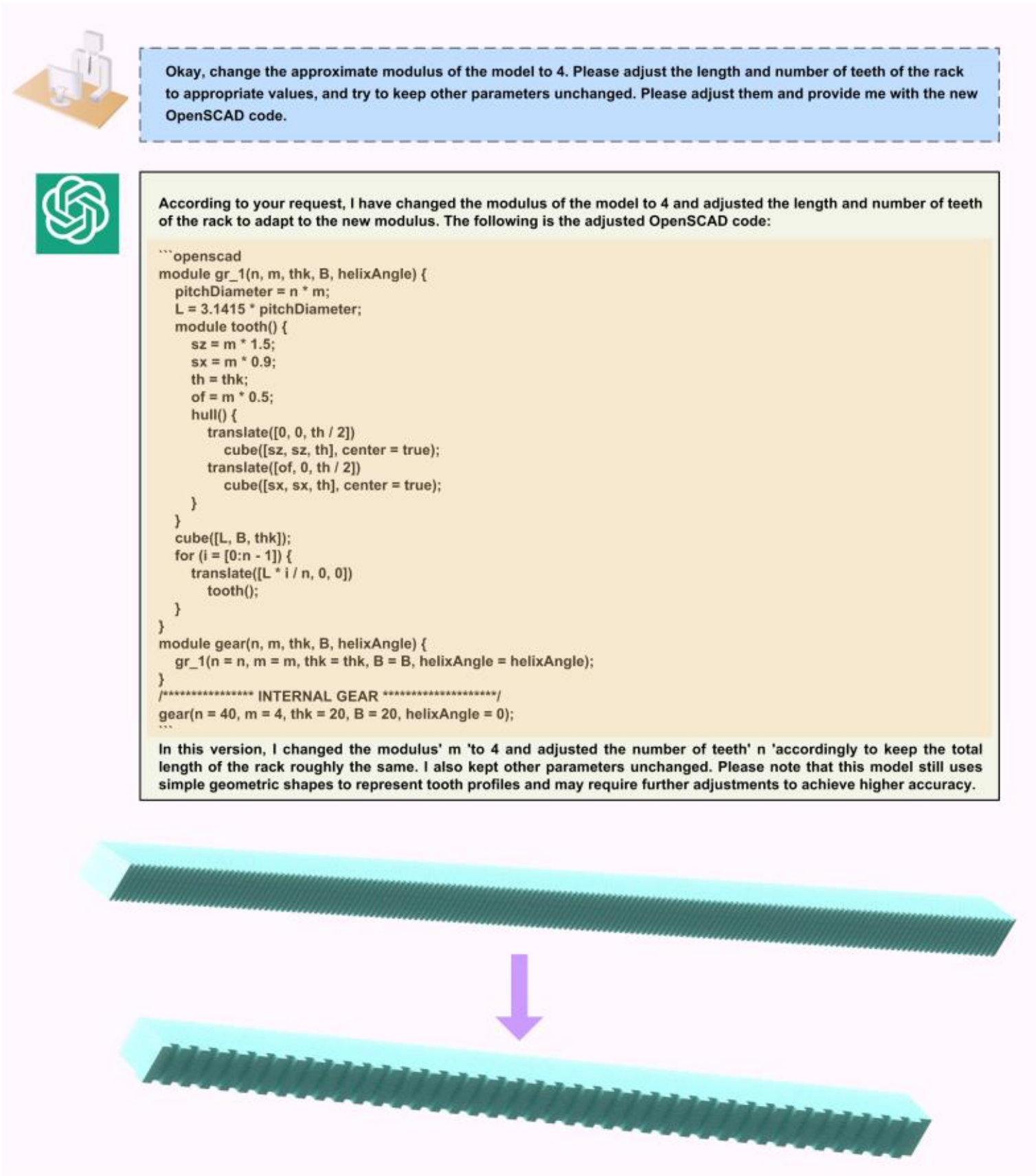

**Figure 10 Spline MDA construction process—Final Evaluation (Example 1).**

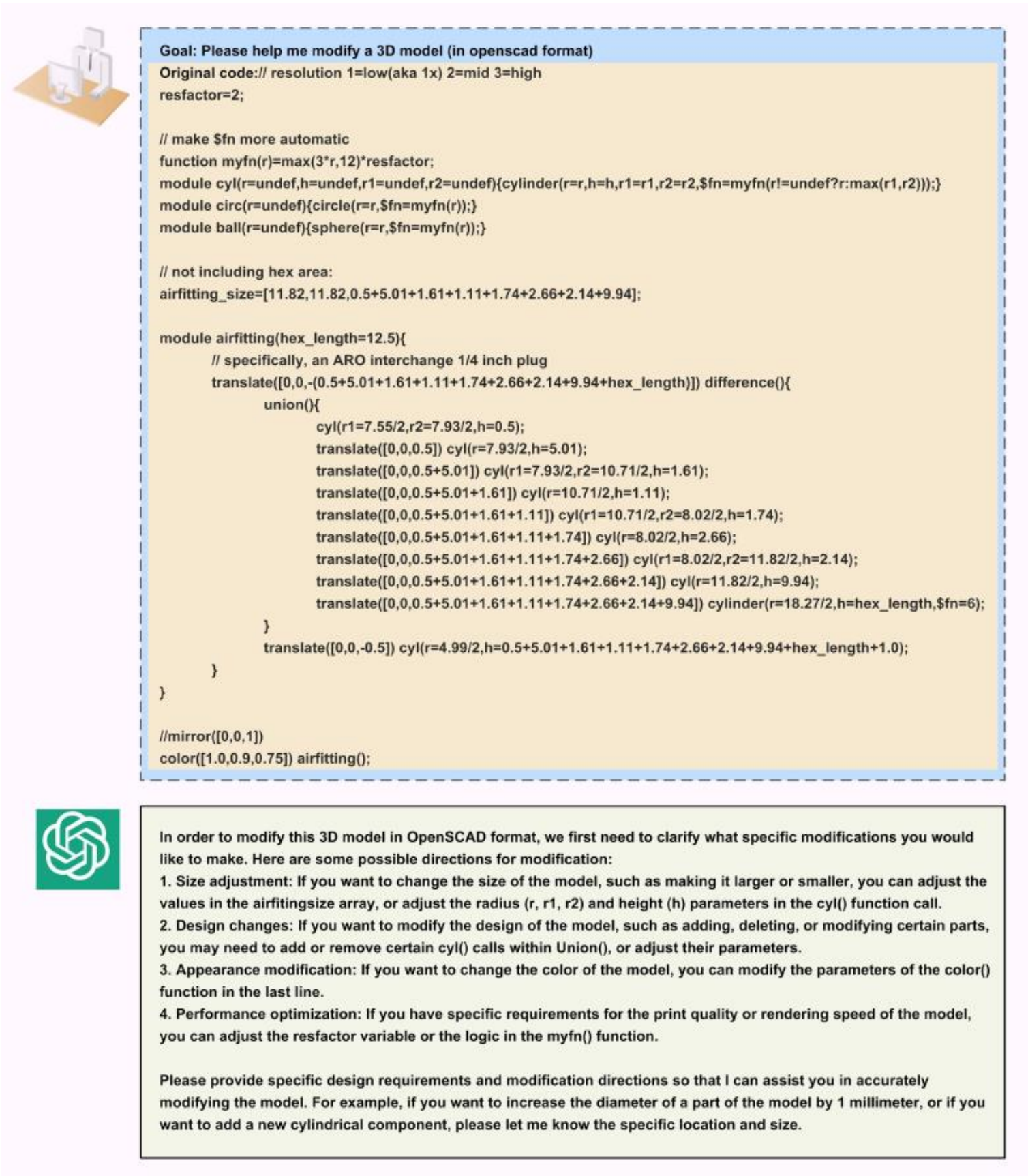

**Figure 11 Joint MDA construction process—Foundational Code Learning (Example 2).**

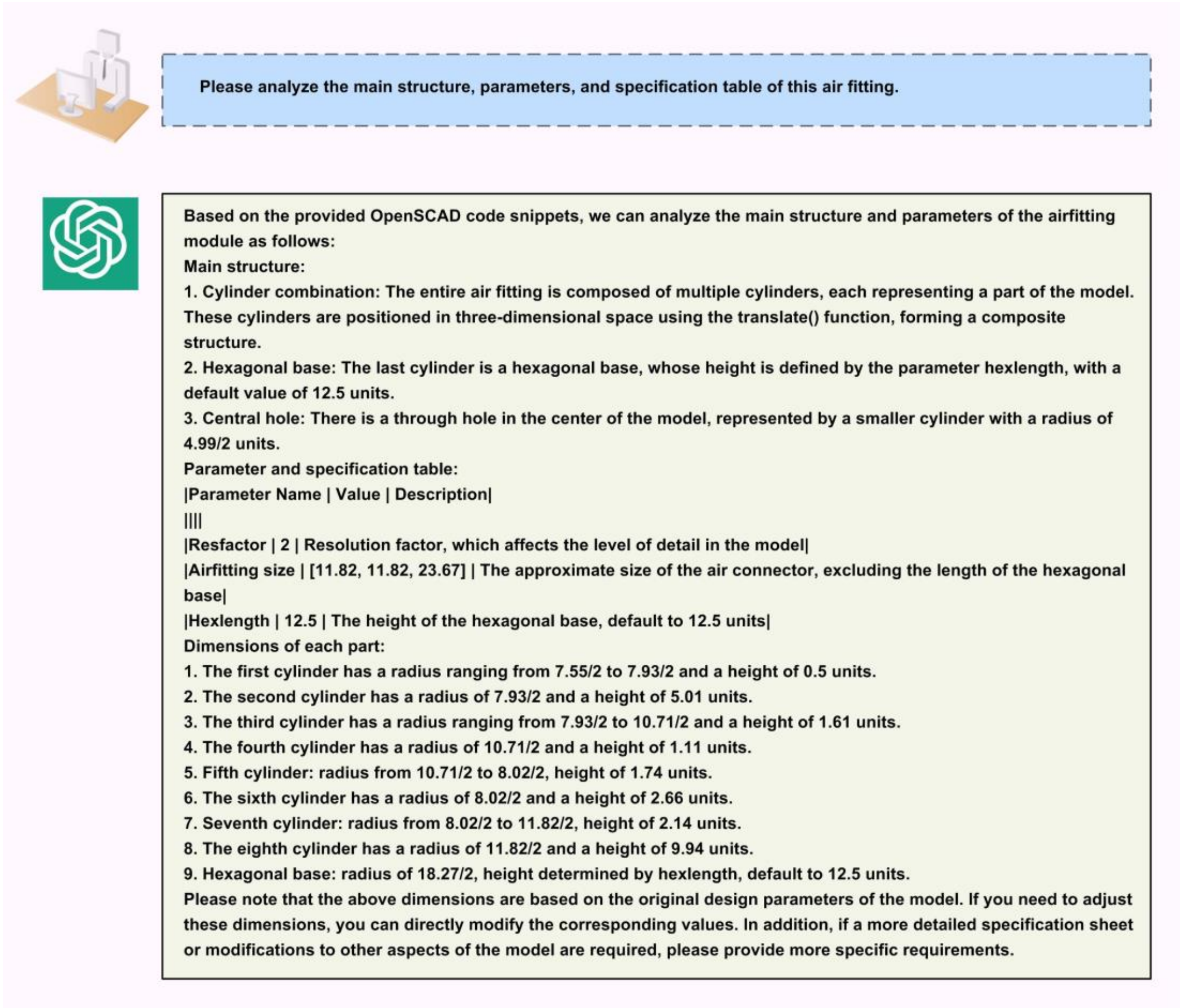

**Figure 12 Joint MDA construction process—Key Feature Extraction (Example 2).**

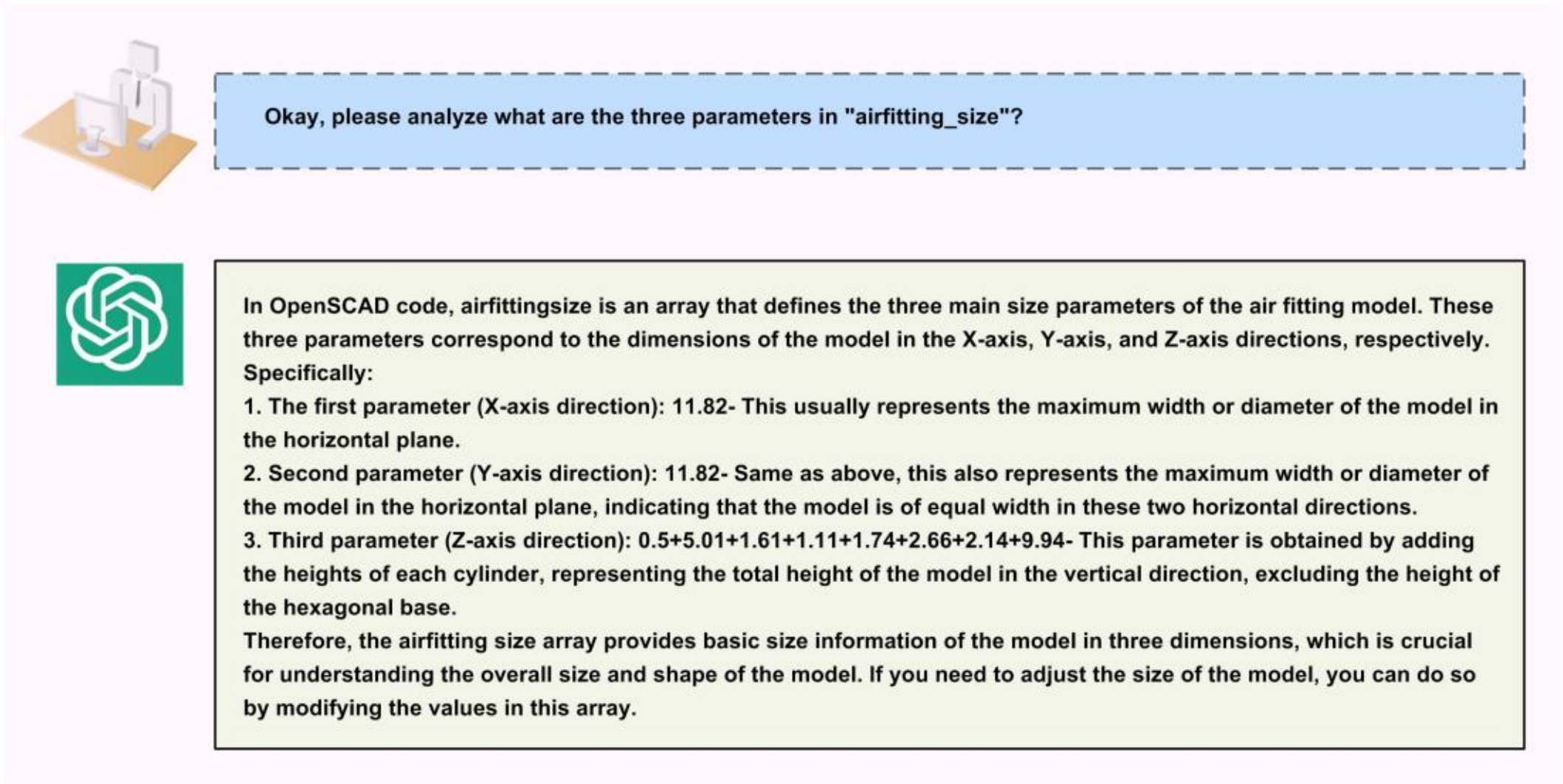

**Figure 13 Joint MDA construction process—Feature Analysis (Example 2).**

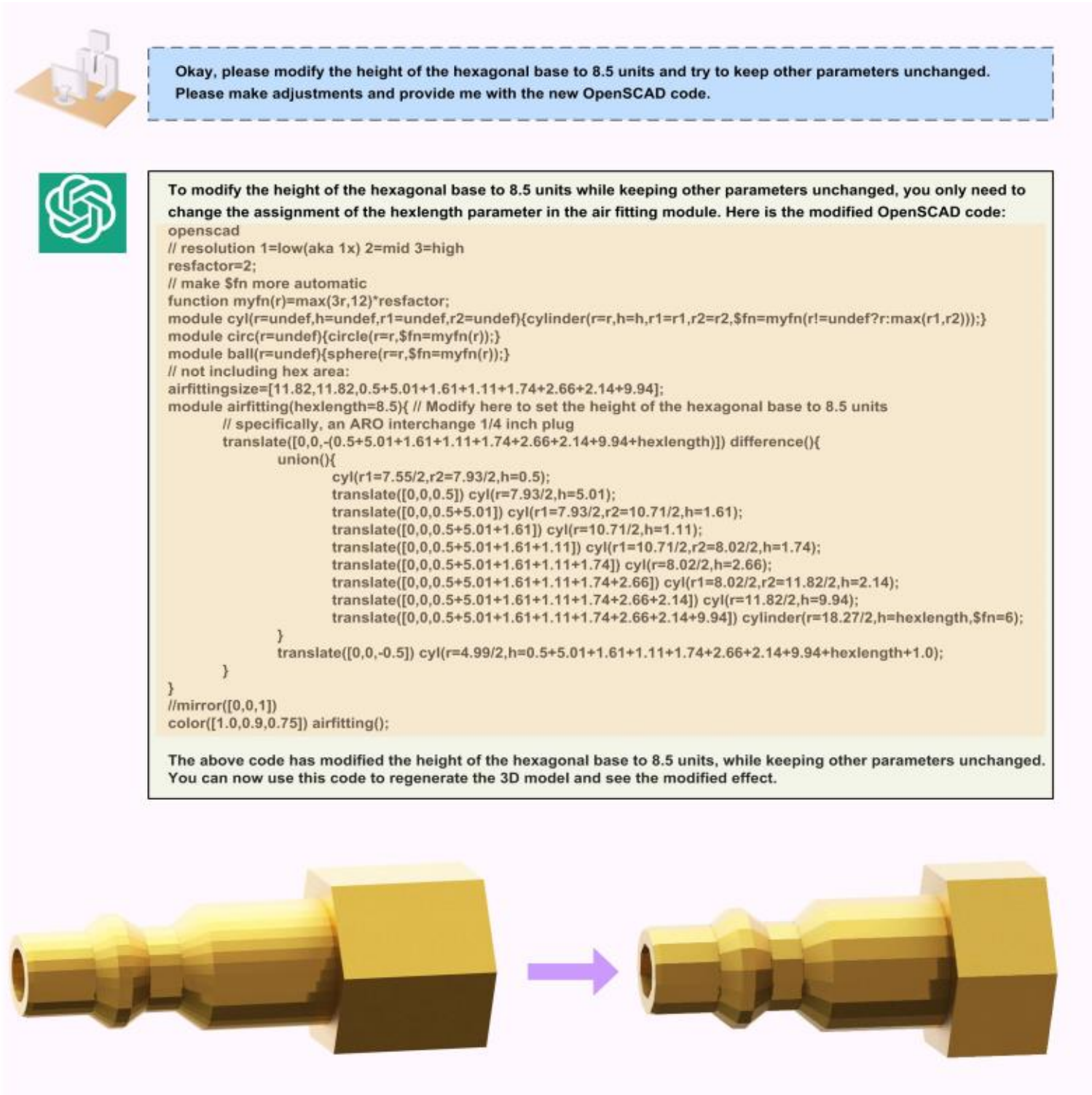

**Figure 14 Joint MDA construction process—First Evaluation (Example 2).**

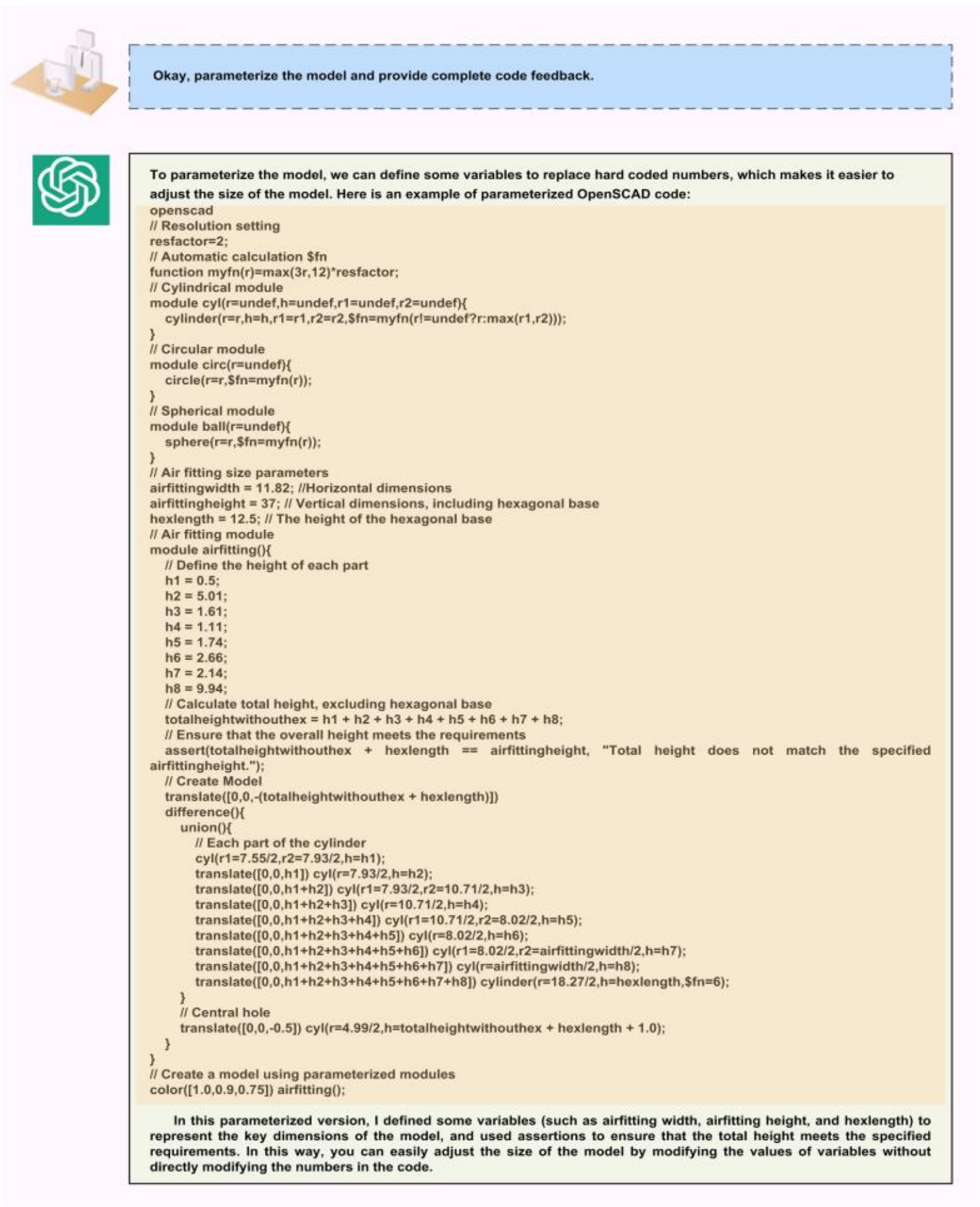

**Figure 15 Joint MDA construction process—Part Parameterization (Example 2).**

Okay, change the total height of the model in the vertical direction (airfitting height) to 41. Please adjust the values of each part h1, h2, h3, h4, h5, h6, h7, and h8 to be appropriate to ensure that 'total_height_without_hex + hex_length == airfitting_height'. Please adjust the other parameters as much as possible and provide me with the new OpenSCAD code.

In order to change the total height of the model in the vertical direction (airfitting height) to 41 units and adjust the heights of each part (h1 to h8), we need to reassign these height values to ensure that 'total_height_without_hex + hex_length == airfitting_height'. Here is a possible adjustment plan:

```
openscad
// Resolution setting
resfactor=2;

// Automatic calculation $fn
function myfn(r)=max(3*r,12)*resfactor;

// Cylindrical module
module cyl(r=undef,h=undef,r1=undef,r2=undef){
   cylinder(r=r,h=h,r1=r1,r2=r2,$fn=myfn(r!=undef?r:max(r1,r2)));
}

// Circular module
module circ(r=undef){
   circle(r=r,$fn=myfn(r));
}

// Spherical module
module ball(r=undef){
   sphere(r=r,$fn=myfn(r));
}

// Air fitting size parameters
airfitting_width = 11.82; // Horizontal dimensions
airfitting_height = 41; // Vertical dimensions, including hexagonal base
hex_length = 12.5; // The height of the hexagonal base

// Air fitting module
module airfitting(){
   // Define the height of each part
   h1 = 0.5;
   h2 = 5.01;
   h3 = 1.61;
   h4 = 1.11;
   h5 = 1.74;
   h6 = 2.66;
   h7 = 2.14;
   h8 = 13.73; // Adjust this value to meet the overall height requirement

   // Calculate total height, excluding hexagonal base
   total_height_without_hex = h1 + h2 + h3 + h4 + h5 + h6 + h7 + h8;

   // Ensure that the overall height meets the requirements    assert(total_height_without_hex + hex_length ==
airfitting_height, "Total height does not match the specified airfitting_height.");

   // Create Model
   translate([0,0,-(total_height_without_hex + hex_length)])
   difference(){
      union(){
         // Each part of the cylinder
         cyl(r1=7.55/2,r2=7.93/2,h=h1);
         translate([0,0,h1]) cyl(r=7.93/2,h=h2);
         translate([0,0,h1+h2]) cyl(r1=7.93/2,r2=10.71/2,h=h3);
         translate([0,0,h1+h2+h3]) cyl(r=10.71/2,h=h4);
         translate([0,0,h1+h2+h3+h4]) cyl(r1=10.71/2,r2=8.02/2,h=h5);
         translate([0,0,h1+h2+h3+h4+h5]) cyl(r=8.02/2,h=h6);
         translate([0,0,h1+h2+h3+h4+h5+h6]) cyl(r1=8.02/2,r2=airfitting_width/2,h=h7);
         translate([0,0,h1+h2+h3+h4+h5+h6+h7]) cyl(r=airfitting_width/2,h=h8);
         translate([0,0,h1+h2+h3+h4+h5+h6+h7+h8]) cylinder(r=18.27/2,h=hex_length,$fn=6);
      }
      // Central hole
      translate([0,0,-0.5]) cyl(r=4.99/2,h=total_height_without_hex + hex_length + 1.0);
   }
}

// Create a model using parameterized modules
color([1.0,0.9,0.75]) airfitting();
```

In this adjustment plan, I retained most of the original height values and only adjusted the value of h8 to ensure that the total height is equal to 41 units. You can further adjust the height of other parts according to your actual needs.

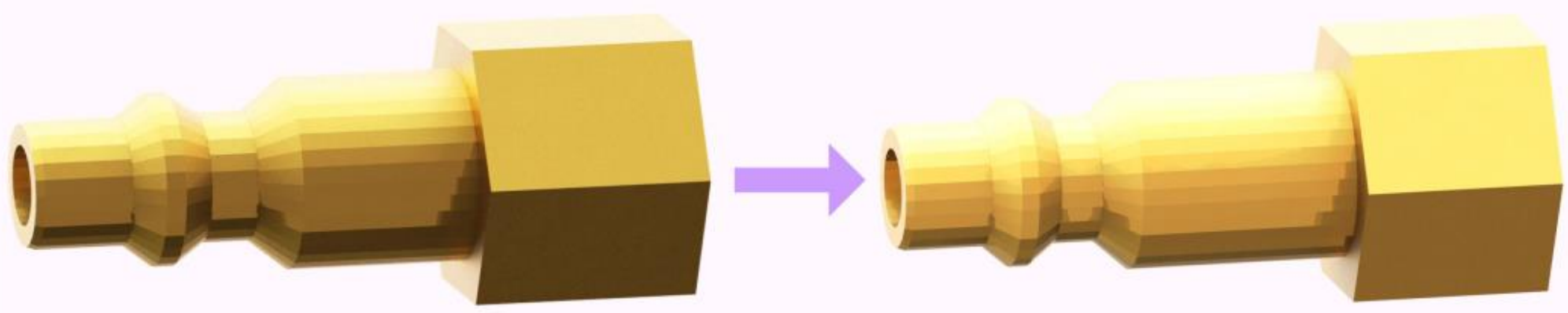

**Figure 16 Joint MDA construction process—Final Evaluation 1 (Example 2).**

Please adjust the diameter of the central hole to 3.7.

To adjust the diameter of the center hole to 3.7 units, you need to modify the radius of the cylinder that created the center hole. In OpenSCAD code, the center hole is defined by the following sections:

```
openscad
translate([0,0,-0.5]) cyl(r=4.99/2,h=totalheightwithouthex + hexlength + 1.0);
```

The r=4.99/2 here defines the radius of the central hole. To adjust its diameter to 3.7 units, you need to set the radius to 3.7/2. The following is the adjusted code snippet:

```
openscad
translate([0,0,-0.5]) cyl(r=3.7/2,h=totalheightwithouthex + hexlength + 1.0);
```

Applying this change to the complete OpenSCAD code will result in the following:

```
openscad
// Resolution setting
resfactor=2;
// Automatic calculation $fn
function myfn(r)=max(3r,12)resfactor;
// Cylindrical module
module cyl(r=undef,h=undef,r1=undef,r2=undef){
   cylinder(r=r,h=h,r1=r1,r2=r2,$fn=myfn(r!=undef?r:max(r1,r2)));
}
// Circular module
module circ(r=undef){
   circle(r=r,$fn=myfn(r));
}
// Spherical module
module ball(r=undef){
   sphere(r=r,$fn=myfn(r));
}
// Air fitting size parameters
airfittingwidth = 11.82; // Horizontal dimensions
airfittingheight = 41; // Vertical dimensions, including hexagonal base
hexlength = 12.5; // The height of the hexagonal base
// Air connector module
module airfitting(){
   // Define the height of each part
   h1 = 0.5;
   h2 = 5.01;
   h3 = 1.61;
   h4 = 1.11;
   h5 = 1.74;
   h6 = 2.66;
   h7 = 2.14;
   h8 = 13.73; // Adjust this value to meet the overall height requirement
   // Calculate total height, excluding hexagonal base
   totalheightwithouthex = h1 + h2 + h3 + h4 + h5 + h6 + h7 + h8;
   // Ensure that the overall height meets the requirements
   assert(totalheightwithouthex + hexlength == airfittingheight, "Total height does not match the specified
airfittingheight.");
   // Create Model
   translate([0,0,-(totalheightwithouthex + hexlength)])
   difference(){
      union(){
         // Each part of the cylinder
         cyl(r1=7.55/2,r2=7.93/2,h=h1);
         translate([0,0,h1]) cyl(r=7.93/2,h=h2);
         translate([0,0,h1+h2]) cyl(r1=7.93/2,r2=10.71/2,h=h3);
         translate([0,0,h1+h2+h3]) cyl(r=10.71/2,h=h4);
         translate([0,0,h1+h2+h3+h4]) cyl(r1=10.71/2,r2=8.02/2,h=h5);
         translate([0,0,h1+h2+h3+h4+h5]) cyl(r=8.02/2,h=h6);
         translate([0,0,h1+h2+h3+h4+h5+h6]) cyl(r1=8.02/2,r2=airfittingwidth/2,h=h7);
         translate([0,0,h1+h2+h3+h4+h5+h6+h7]) cyl(r=airfittingwidth/2,h=h8);
         translate([0,0,h1+h2+h3+h4+h5+h6+h7+h8]) cylinder(r=18.27/2,h=hexlength,$fn=6);
      }
      // Central hole
      translate([0,0,-0.5]) cyl(r=3.7/2,h=totalheightwithouthex + hexlength + 1.0);
   }
}
// Create a model using parameterized modules
color([1.0,0.9,0.75]) airfitting();
```

Now, the diameter of the central hole has been successfully adjusted to 3.7 units.

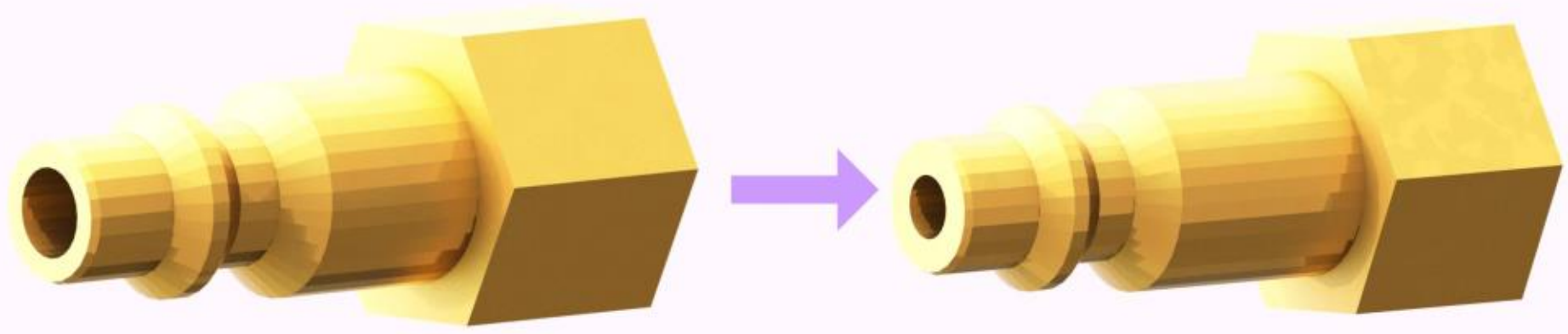

**Figure 17 Joint MDA construction process—Final Evaluation 2 (Example 2).**

## 5. Conclusion

In conclusion, we find that the application of large language models to 3D mechanical design holds great potential, and we present our experimental results. While directly instructing an LLM to complete the task is not very effective, our theory-guided approach to constructing a Mechanical Design Agent (MDA) has led to LLMs demonstrating modeling abilities comparable to those of professional designers. The results are promising. However, it must be acknowledged that there is still considerable work needed to completely replace designers in the field. MDAs still require extensive learning and guidance to become experts in all areas of mechanical design, rather than excelling in only specific parts.

For the field of mechanical design, our work provides valuable practical methods for other researchers. Although we have made substantial progress, the potential to fully leverage LLMs in mechanical design remains vast, presenting both opportunities and challenges.

## 6. Acknowledgment

This research was funded by National Natural Science Foundation of China:52205245.

## 7. References

[1] R. Bommasani *et al.*, “On the Opportunities and Risks of Foundation Models,” Jul. 12, 2022, *arXiv*: arXiv:2108.07258. doi: 10.48550/arXiv.2108.07258.

[2] V. Sanh, L. Debut, J. Chaumond, and T. Wolf, “DistilBERT, a distilled version of BERT: smaller, faster, cheaper and lighter,” Feb. 29, 2020, *arXiv*: arXiv:1910.01108. doi: 10.48550/arXiv.1910.01108.

[3] R. Schwartz, J. Dodge, N. A. Smith, and O. Etzioni, “Green AI,” *Commun. ACM*, vol. 63, no. 12, pp. 54–63, Nov. 2020, doi: 10.1145/3381831.

[4] E. Alsentzer *et al.*, “Publicly Available Clinical BERT Embeddings,” Jun. 20, 2019, *arXiv*: arXiv:1904.03323. doi: 10.48550/arXiv.1904.03323.

[5] K. Huang, J. Altosaar, and R. Ranganath, “ClinicalBERT: Modeling Clinical Notes and Predicting Hospital Readmission,” Nov. 28, 2020, *arXiv*: arXiv:1904.05342. doi: 10.48550/arXiv.1904.05342.

[6] A. Zhavoronkov *et al.*, “Deep learning enables rapid identification of potent DDR1 kinase inhibitors,” *Nat Biotechnol*, vol. 37, no. 9, pp. 1038–1040, Sep. 2019, doi: 10.1038/s41587-019-0224-x.

[7] K. Ethayarajh and D. Jurafsky, “Utility is in the Eye of the User: A Critique of NLP Leaderboards,” Mar. 03, 2021, *arXiv*: arXiv:2009.13888. doi: 10.48550/arXiv.2009.13888.

[8] E. M. Bender, T. Gebru, A. McMillan-Major, and S. Shmitchell, “On the Dangers of Stochastic Parrots: Can Language Models Be Too Big? 🦜,” in *Proceedings of the 2021 ACM Conference on Fairness, Accountability, and Transparency*, in FAccT ’21. New York, NY, USA: Association for Computing Machinery, Mar. 2021,

pp. 610–623. doi: 10.1145/3442188.3445922.
[9] K. D. D. Willis *et al.*, “Fusion 360 gallery: a dataset and environment for programmatic CAD construction from human design sequences,” *ACM Trans. Graph.*, vol. 40, no. 4, p. 54:1-54:24, Jul. 2021, doi: 10.1145/3450626.3459818.
[10] “FeatureScript introduction.” Accessed: Aug. 03, 2024. [Online]. Available: https://cad.onshape.com/FsDoc/
[11] P. Prusinkiewicz and A. Lindenmayer, *The Algorithmic Beauty of Plants*. Springer Science & Business Media, 2012.
[12] Y. Zhang, M. Yang, R. Baghdadi, S. Kamil, J. Shun, and S. Amarasinghe, “GraphIt: a high-performance graph DSL,” *Proc. ACM Program. Lang.*, vol. 2, no. OOPSLA, p. 121:1-121:30, Oct. 2018, doi: 10.1145/3276491.
[13] T. Du *et al.*, “InverseCSG: automatic conversion of 3D models to CSG trees,” *ACM Trans. Graph.*, vol. 37, no. 6, p. 213:1-213:16, Dec. 2018, doi: 10.1145/3272127.3275006.
[14] A. Zhao *et al.*, “RoboGrammar: graph grammar for terrain-optimized robot design,” *ACM Trans. Graph.*, vol. 39, no. 6, p. 188:1-188:16, Nov. 2020, doi: 10.1145/3414685.3417831.
[15] L. Makatura *et al.*, “Procedural Metamaterials: A Unified Procedural Graph for Metamaterial Design,” *ACM Trans. Graph.*, vol. 42, no. 5, p. 168:1-168:19, Jul. 2023, doi: 10.1145/3605389.
[16] P. Müller, P. Wonka, S. Haegler, A. Ulmer, and L. Van Gool, “Procedural modeling of buildings,” in *ACM SIGGRAPH 2006 Papers*, in SIGGRAPH ’06. New York, NY, USA: Association for Computing Machinery, Jul. 2006, pp. 614–623. doi: 10.1145/1179352.1141931.
[17] M. Guo, V. Thost, B. Li, P. Das, J. Chen, and W. Matusik, “Data-Efficient Graph Grammar Learning for Molecular Generation,” Mar. 15, 2022, *arXiv*: arXiv:2203.08031. doi: 10.48550/arXiv.2203.08031.
[18] A. Vaswani *et al.*, “Attention is All you Need,” in *Advances in Neural Information Processing Systems*, Curran Associates, Inc., 2017. Accessed: Aug. 03, 2024. [Online]. Available: https://proceedings.neurips.cc/paper_files/paper/2017/hash/3f5ee243547dee91fbd053c1c4a845aa-Abstract.html
[19] Y. Ganin, S. Bartunov, Y. Li, E. Keller, and S. Saliceti, “Computer-Aided Design as Language,” in *Advances in Neural Information Processing Systems*, Curran Associates, Inc., 2021, pp. 5885–5897. Accessed: Jun. 12, 2024. [Online]. Available: https://proceedings.neurips.cc/paper/2021/hash/2e92962c0b6996add9517e4242ea9bdc-Abstract.html
[20] C. Li, H. Pan, A. Bousseau, and N. J. Mitra, “Free2CAD: parsing freehand drawings into CAD commands,” *ACM Trans. Graph.*, vol. 41, no. 4, p. 93:1-93:16, Jul. 2022, doi: 10.1145/3528223.3530133.
[21] A. Seff, W. Zhou, N. Richardson, and R. P. Adams, “Vitruvion: A Generative Model of Parametric CAD Sketches,” Apr. 28, 2022, *arXiv*: arXiv:2109.14124. doi: 10.48550/arXiv.2109.14124.
[22] X. Xu, K. D. D. Willis, J. G. Lambourne, C.-Y. Cheng, P. K. Jayaraman, and Y.

Furukawa, “SkexGen: Autoregressive Generation of CAD Construction Sequences with Disentangled Codebooks,” Jul. 11, 2022, *arXiv*: arXiv:2207.04632. doi: 10.48550/arXiv.2207.04632.

[23] X. Xu, P. K. Jayaraman, J. G. Lambourne, K. D. D. Willis, and Y. Furukawa, “Hierarchical Neural Coding for Controllable CAD Model Generation,” Jun. 30, 2023, *arXiv*: arXiv:2307.00149. doi: 10.48550/arXiv.2307.00149.

[24] S. Zhou, T. Tang, and B. Zhou, “CADParser: A Learning Approach of Sequence Modeling for B-Rep CAD,” presented at the Thirty-Second International Joint Conference on Artificial Intelligence, Aug. 2023, pp. 1804–1812. doi: 10.24963/ijcai.2023/200.

[25] J. G. Lambourne, K. D. D. Willis, P. K. Jayaraman, A. Sanghi, P. Meltzer, and H. Shayani, “BRepNet: A Topological Message Passing System for Solid Models,” presented at the Proceedings of the IEEE/CVF Conference on Computer Vision and Pattern Recognition, 2021, pp. 12773–12782. Accessed: Aug. 03, 2024. [Online]. Available: https://openaccess.thecvf.com/content/CVPR2021/html/Lambourne_BRepNet_A_Topological_Message_Passing_System_for_Solid_Models_CVPR_2021_paper.html

[26] R. Wu, C. Xiao, and C. Zheng, “DeepCAD: A Deep Generative Network for Computer-Aided Design Models,” Aug. 15, 2021, *arXiv*: arXiv:2105.09492. Accessed: Apr. 09, 2024. [Online]. Available: http://arxiv.org/abs/2105.09492

[27] W. Ma, M. Xu, X. Li, and X. Zhou, “MultiCAD: Contrastive Representation Learning for Multi-modal 3D Computer-Aided Design Models,” in *Proceedings of the 32nd ACM International Conference on Information and Knowledge Management*, in CIKM ’23. New York, NY, USA: Association for Computing Machinery, Oct. 2023, pp. 1766–1776. doi: 10.1145/3583780.3614982.

[28] “ChatGLM.” Accessed: Aug. 05, 2024. [Online]. Available: https://chatglm.cn/

[29] “OpenSCAD - The Programmers Solid 3D CAD Modeller.” Accessed: Aug. 05, 2024. [Online]. Available: https://openscad.org/